\documentclass[twocolumn,amsmath,amssymb]{revtex4}

\usepackage{graphicx}
\usepackage{multirow}

\usepackage{color}

\usepackage{amsmath}
\usepackage{epsfig}
\usepackage{amsfonts}
\usepackage{amssymb}
\usepackage{cancel}

\begin{document}

\title{Probing top-Higgs non-standard interactions at the LHC}

\author{C. Degrande$^{a,b}$, J.-M. G\'erard$^b$, C. Grojean$^{c}$, F. Maltoni$^b$, G. Servant$^{c,d}$}

\affiliation{$^a$Department of Physics, University of Illinois at Urbana-Champaign, 1110 W. Green Street, Urbana, IL 61801\\ 
$^b$Centre for Cosmology, Particle Physics and Phenomenology (CP3)\\
Chemin du Cyclotron 2, Universit\'e catholique de Louvain, Belgium,\\
$^c$CERN Physics Department, Theory Division, CH-1211 
Geneva 23, Switzerland\\
$^d$Institut de Physique Th\'eorique, CEA/Saclay, F-91191 
Gif-sur-Yvette C\'edex, France
}

\begin{abstract}

Effective interactions involving both the top quark and the Higgs field  Êare among the least  constrained of all possible (gauge invariant) dimension-six operators in the  Standard Model.Ê Such a handful of operators, in particular ÊtheÊ top quark chromomagneticÊ dipole moment, might encapsulate signs of the new physics responsible for electroweakÊ symmetry breaking. In this work, we compute the contributions of these operators to inclusive Higgs and $t \bar t h$ production. We argue that: Êi) rather strong constraints on the overall size of these operatorsÊ 
can   already   be  obtained from the current limits/evidence on Higgs production at the LHC; ii) Ê$t \bar t h$ production will provideÊ further key information that is complementary to $t\bar t$ measurements, and the possibility of discriminating among different contributions by performing accurate measurements of total and 
differential rates.

\end{abstract}

\maketitle

\section{Introduction}

The Standard Model (SM) has been tested with an impressive accuracy and is so far in excellent agreement with the experimental data. The room left for new physics at the TeV scale is therefore getting more and more squeezed, thanks to the LHC. Effective field theory (EFT) provides a model independent parametrization of the potential deviations from the SM while keeping its successes if the new degrees of freedom are heavy.  EFT has been intensively used for instance  in flavor physics to translate the accuracy of the measurements into strong constraints on the coefficients of the associated operators~\cite{Bona:2007vi}. Slightly softer constraints on the operators involving weak bosons have also been derived from the electroweak precision measurements~\cite{Han:2004az,Barbieri:2004qk}. In comparison, the operators involving the top quark are poorly constrained so far~\cite{Zhang:2012cd}, especially the chromomagnetic moment operator of the top quark~\cite{Degrande:2010kt,Kamenik:2011dk},  while those involving the Higgs field remain largely unconstrained.
However, this status is about to change. In particular, modifications of the top quark interactions can significantly change the main Higgs production mechanism
 at hadron colliders, 
 which is under scrutiny at the LHC.

In this paper, we focus on operators that involve both
the top quark and the Higgs field. Not only they are little
tested, but  it is also where one might expect new 
physics associated with electroweak symmetry breaking to show up.
First, we compute their contributions  to  $gg\to h$  due to a top loop. 
The only non-trivial contribution due to the chromomagnetic operator
is logarithmic divergent and can be as large as the SM one. We then derive the 
 constraints  from the experimental bound on the Higgs production rate. 
 Higgs production by gluon fusion alone does not allow to distinguish the new contributions since they are all proportional to the SM amplitude. In 
 Section~\ref{tth}, we argue that $t\bar{t}h$ production can provide complementary information
  to further constrain and to disentangle the various top-Higgs operators.

\section{Operators of interest}

Recently, constraints on  effective Higgs interactions from the latest Higgs searches have been derived 
\cite{Carmi:2012yp,Azatov:2012bz,Espinosa:2012ir,Giardino:2012ww,Ellis:2012rx,Azatov:2012rd,Farina:2012ea},
with an emphasis on the $d=6$ operators built from the Higgs  and SM gauge bosons.
These papers display global fits in a large parameter space.
While Ref.~\cite{Azatov:2012bz} and 
Ref.~\cite{Espinosa:2012ir},
 are restricted to a particular UV set-up where only a sub-class of operators are important, Refs.~\cite{Carmi:2012yp,Giardino:2012ww} 
 included the modification of all Higgs interactions to the
SM particles but considered that only the Yukawa coupling of the
fermions were changed, and therefore have not considered the chromomagnetic operator.
The spirit of this work is different in that our motivation is to focus only on $d=6$  operators that involve both the Higgs 
field and the top quark. We study their effect on Higgs production by gluon fusion and associated with  a $t\bar{t}$ pair, assuming in particular that $hWW$ and $hZZ$ tree-level couplings are not affected by new physics. 
The results of our analysis can easily be updated once the $hWW$ and $hZZ$ couplings are better determined.
We start with the effective lagrangian  \cite{Buchmuller:1985jz,Grzadkowski:2010es,Buchalla:2012qq}
\begin{equation}
 \mathcal{L}=\mathcal{L}^{SM}+\sum \frac{c_i}{\Lambda^2} \mathcal{O}_{i} + {O}\left( \frac{1}{\Lambda^4} \right).
\end{equation}
The chromomagnetic dipole moment operator modifies the interactions between the gluons and the top quark, 
\begin{equation}
\mathcal{O}_{hg}=\left(\bar{Q}_L H\right)\sigma^{\mu\nu}T^at_R G_{\mu\nu}^a,\label{chromo}
\end{equation}
where $\sigma^{\mu\nu}=\frac{i}{2}\left[\gamma^\mu,\gamma^\nu\right]$ and $T^a$ is such that $\text{Tr}\left(T^aT^b\right)=\delta^{ab}/2$. 
Besides, one operator contains the top density
\begin{equation} 
{\mathcal O}_{Hy}  =  H^\dagger H \left( H \bar{Q}_L \right) t_R
\label{htQop}
\end{equation}
and three operators can be built from the top and Higgs currents,
\begin{eqnarray}
{\mathcal O}_{Ht} & = & H^\dagger D_\mu H \bar{t}_R \gamma^\mu t_R\nonumber\\
{\mathcal O}_{HQ} & = & H^\dagger D_\mu H \bar{Q}_L \gamma^\mu Q_L\nonumber\\
{\mathcal O}^{(3)}_{HQ} & = & H^\dagger \sigma^I D_\mu H \bar{Q}_L \sigma^I \gamma^\mu Q_L \  .    \label{htop} 
\end{eqnarray}
Other operators of dimension 6 play a role in the top-Higgs interaction even though
they do not contain both fields. One of them is 
\begin{equation}  
{\mathcal O}_{H}  =  \partial_\mu \left(H^\dagger H\right) \partial^\mu \left(H^\dagger H\right),\label{hop}
\end{equation}
which amounts to an overall renormalization of the Higgs wave function and
therefore to a trivial shift of the top-quark Yukawa coupling~\cite{Giudice:2007fh}.  

\begin{figure}[t]
	\centering
		\includegraphics[width=0.18\textwidth]{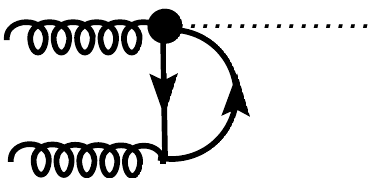}
		\includegraphics[width=0.18\textwidth]{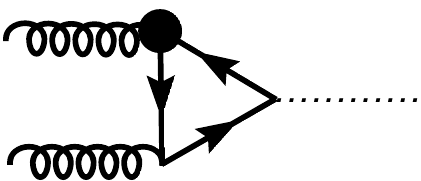}
		\includegraphics[width=0.18\textwidth]{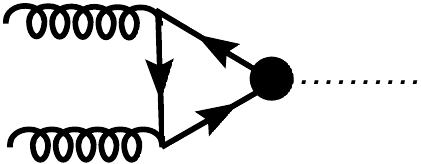}
	\caption{$gg\to h$ production. The first two diagrams are the contributions to $O_{HG}$ from $O_{hg}$. The third one
	 is induced by $O_{Hy}$ and $O_{H}$. The operators of Eq.~(\ref{htop}) do 
	 not contribute to $\mathcal{O}_{HG}$ (see Sec.~\ref{ggh}).}
\label{fig:effectiveOHG}
\end{figure}

The corrections from those operators to Higgs production by gluon fusion are shown in Fig.~\ref{fig:effectiveOHG}.
 In the large top mass limit, the contribution of the operators in Eqs.~(\ref{chromo},\ref{htQop}) can be seen as corrections to the $\mathcal{O}_{HG}$ operator 
\begin{equation}
\mathcal{O}_{HG}=\frac{1}{2}H^\dagger H G_{\mu\nu}^{a}G^{\mu\nu}_a
\end{equation} 
generated by the scale anomaly.
Therefore, we are going to 
derive the constraints on  $\mathcal{O}_{HG}$ from Higgs production,  which we will then re-express in terms of limits on a combination of the above operators.

One should remark that not only the Higgs production rate is sensitive to the modifications of the top interactions but also the $h \to \gamma \gamma$ decay. 
The operator $O_H$ does not change
the branching ratios since it multiplies all partial widths by the same
factor. However, $\mathcal{O}_{Hy}$
and the electromagnetic version of $\mathcal{O}_{hg}$ induce 
\begin{equation}
\mathcal{O}_{H\gamma}=\frac{1}{2}H^\dagger H F_{\mu\nu}F^{\mu\nu}.
\end{equation}
The main effect of this operator will be to relax the constraints from the $h \to \gamma \gamma$  channel.
We reiterate that we do not consider corrections to $hWW$ and $hZZ$ vertices.
New top interactions affect all these channels at one-loop. However,
their effects to the loop-induced processes $ h \rightarrow \gamma \gamma$ and $gg \to h$
are expected to be relatively larger than for $h \to WW$  and $h \to ZZ$ because
the new operators modify the SM leading order in the first case and the NLO
corrections in the second.

\section{Higgs production by gluon fusion}\label{ggh}

\begin{figure}[tb]
	\centering
		\includegraphics[width=0.45\textwidth]{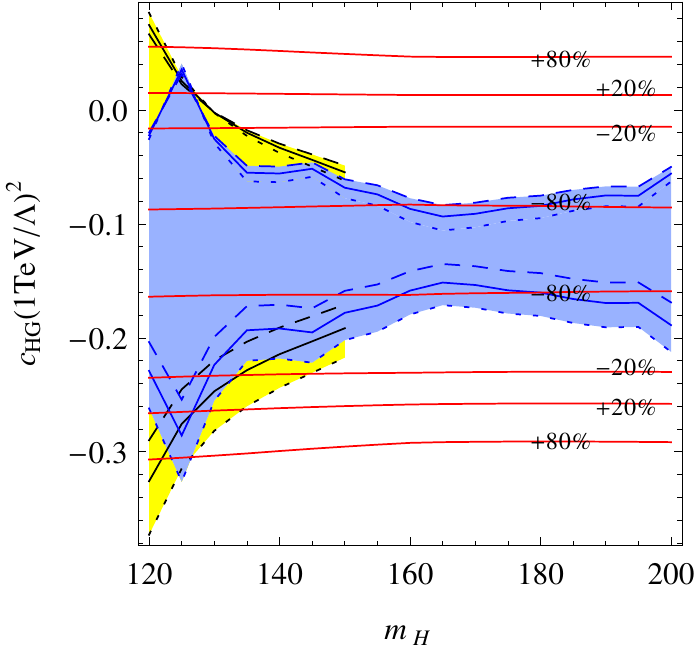}
	\caption{Region allowed at 95\% C.L. by the ATLAS upper bound on the Higgs production cross-section~\cite{ATLAS:2012ae}  for $\mu_R=\mu_F={m_H}/{2}$ (solid line). The errors are estimated by varying the renormalization and factorization scales from $\mu_R=\mu_F={m_H}/{4}$ (dotted line) to $\mu_R=\mu_F=m_H$ (dashed line). The blue region uses the combination of all channels. The yellow region is obtained using the strongest constraint among the $WW$ and $ZZ$ channels. The red lines show the relative deviation compared to the SM Higgs production rate. }
	\label{fig:cHGconst}
\end{figure}

$\mathcal{O}_{HG}$ is the only dimension-six operator inducing Higgs production by gluon fusion at tree-level.
Its effect on the partonic cross-section is  (see also Refs.~\cite{Manohar:2006gz,Pierce:2006dh})
\begin{equation}
\sigma\left(gg\to h\right)=\sigma\left(gg\to h\right)_{SM}\left(1+ \frac{c_{HG}}{\Lambda^2}\frac{6\pi v^2}{\alpha_s}\right)^2 \ ,
\end{equation} 
where we have taken the heavy top limit for the SM, i.e., $m_t > m_H/2$, and $v\approx 246$~GeV is the Higgs vacuum expectation value (vev).
The contribution from $\mathcal{O}_{HG}$  is quite large compared to the SM one ($6\pi v^2/\alpha_s\sim10$ TeV$^2$) because the latter is only generated at the loop-level. 
Consequently, the upper limits on the Higgs production cross-section from the Tevatron~\cite{Aaltonen:2011gs} and the LHC~\cite{ATLAS:2011h,CMS:2011h, ATLAS:2012ae} strongly constrain the allowed range for $c_{HG}$, as shown on Fig.~\ref{fig:cHGconst}. 
\begin{figure}[t]
	\centering
		\includegraphics[width=0.40\textwidth]{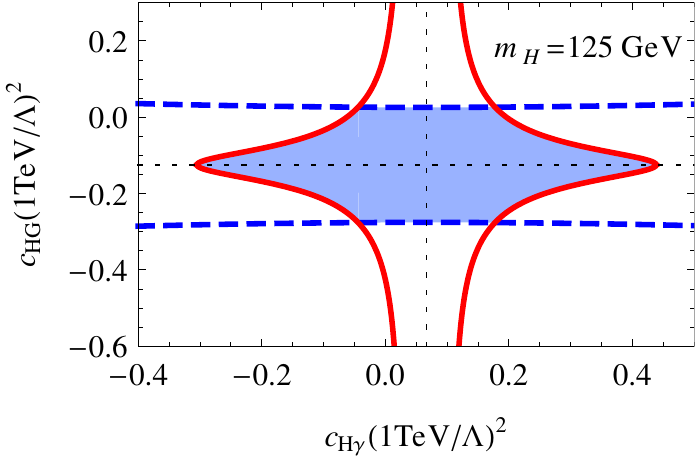}
	\caption{The dashed blue and solid red lines are  the limits from $h\to (WW,ZZ)$  and $h\to \gamma \gamma$ respectively.
 The $WW/ZZ$ constraints on $c_{HG}$ are 
 stronger only when the branching ratio to $\gamma \gamma$ goes below $10^{-3}$ (SM value), corresponding to $0 \lesssim c_{H\gamma}\lesssim 0.1$. For larger branching ratio, the
$\gamma \gamma$ constraints are stronger and do not allow for large values of $c_{H\gamma}$. Note that the allowed region is symmetric along the dotted black lines  where $\sigma (gg \to h)=0$ and $\Gamma (h \to \gamma \gamma)=0$. We have checked that a more refined analysis combining all the channels along the lines of Ref.~\cite{Espinosa:2012ir} gives qualitatively similar results, although slightly more constraining of course.
}
	\label{fig:cHgamma}
\end{figure}
For this figure, we assume that only $\mathcal{O}_{HG}$ is added to the SM Lagrangian, i.e., we neglect the modifications of the other production mechanisms or of the decay widths except for $h\to gg$. We used the same NNLO K factor for the contribution of the $\mathcal{O}_{HG}$ as for the SM~\cite{Dittmaier:2011ti} since both amplitudes are the same up to a global factor. The errors on these limits have been estimated by varying simultaneously the renormalization ($\mu_R$) and factorization scales ($\mu_F$) for the SM and the $\mathcal{O}_{HG}$ tree-level contributions. Other theoretical errors are much smaller.
For $m_H=125$ GeV, we obtain $-0.29 \lesssim c_{HG} (\mbox{TeV}^2/\Lambda^2)\lesssim 0.036$. We also show in yellow how the constraints on $c_{HG}$ are relaxed when including the effect of  $\mathcal{O}_{H\gamma}$.
The exclusion in the plane $(c_{HG},c_{H\gamma})$ is shown in Fig.~\ref{fig:cHgamma}.
Again,  Fig.~\ref{fig:cHGconst} is valid only for SM $hWW$ and $hZZ$ couplings but a similar plot can be drawn once the actual values of $hWW$ and $hZZ$ will be determined.

The constraints on $c_{HG}$ of Fig.~\ref{fig:cHGconst} translate into  constraints on a combination of the coefficients of the operators
Eqs.~(\ref{chromo})$-$(\ref{hop}). Contrary to the result reported in the first version of our paper as well as in Ref.~\cite{Choudhury:2012uv}, the one-loop correction from $\mathcal{O}_{hg}$ to the operator $\mathcal{O}_{HG}$ diverges logarithmically in a way consistent with the general expectations from
dimension-six operators. Its one-loop contribution can be written as
\begin{equation}
\delta c_{HG}=\frac{g_sy_t}{4\pi^2}\Re{c_{hg}}\log\left(\frac{\Lambda^2}{m_t^2}\right).
\label{eq:deltachg}
\end{equation}
%
The operators ${\mathcal O}_{Hy}$ and ${\mathcal O}_{H}$ renormalize the top mass 
\begin{equation}
m_t = y_t \frac{v}{\sqrt 2} - \frac{\Re\left(c_{Hy}\right)}{2\sqrt{2}}\frac{v^3}{\Lambda^2}
\end{equation}
and/or the top Yukawa coupling,
\begin{eqnarray}
\mathcal{L}^{ht\bar{t}}&=&\bar{t} t \frac{h}{\sqrt2} \left( y_t - \left(\frac{3}{2}\Re\left(c_{Hy} \right) +y_t  c_H\right) \frac{v^2}{\Lambda^2}\right) \nonumber\\
 &=&\bar{t} t h \frac{m_t}{v}\left( 1 - c_y\frac{v^2}{\Lambda^2}\right) \ ,
 \label{eq:httren}
\end{eqnarray}
where
\begin{equation}
 c_y=c_H+\frac{v}{\sqrt2 m_t}\Re\left(c_{Hy}\right)\ .
 \label{eq:cy}
 \end{equation}
  Their contributions are then easily obtained as a simple rescaling of the SM contribution~\cite{Low:2010mr,Giudice:2007fh}
 \begin{equation}
\frac{\delta c_{HG}}{\Lambda^2}=\frac{\alpha_s}{6\pi v^2} \times (-c_y\frac{v^2}{\Lambda^2}).
\end{equation} 
The other three operators listed in Eq.~(\ref{htop}) do not contribute to  Higgs production by gluon fusion. In fact, the vertex $ht\bar{t}$ comes from the sum of those operators and of their Hermitian conjugates~\footnote{This combination is invariant under custodial symmetry and can thus not be constrained by the $\rho$ parameter.}. The relevant part of the operators can thus be written as 
\begin{eqnarray}
\partial_\mu \left(H^\dagger H \right)\bar{t} \gamma^\mu \gamma_{\pm} t 
&\propto& \left(H^\dagger H \right)\partial_\mu\frac{(J^\mu\pm J^\mu_5)}{2} \nonumber\\
&\propto& \left(H^\dagger H \right)\partial_\mu J^\mu_5
\end{eqnarray}
because the vector current is conserved. Their contributions to Higgs production through the effective operator $H^\dagger H G^{\mu\nu}\widetilde{G}_{\mu\nu}$, generated by the axial anomaly, vanish in the SM due to parity.   This result  is consistent with the operator relations derived in Ref.~\cite{AguilarSaavedra:2009mx}. In Two-Higgs-Doublet-Models with a light pseudo-scalar, this effective operator should be taken into account \cite{Cervero:2012cx}.

\begin{figure}[tb]
\centering
		\includegraphics[width=0.3\textwidth]{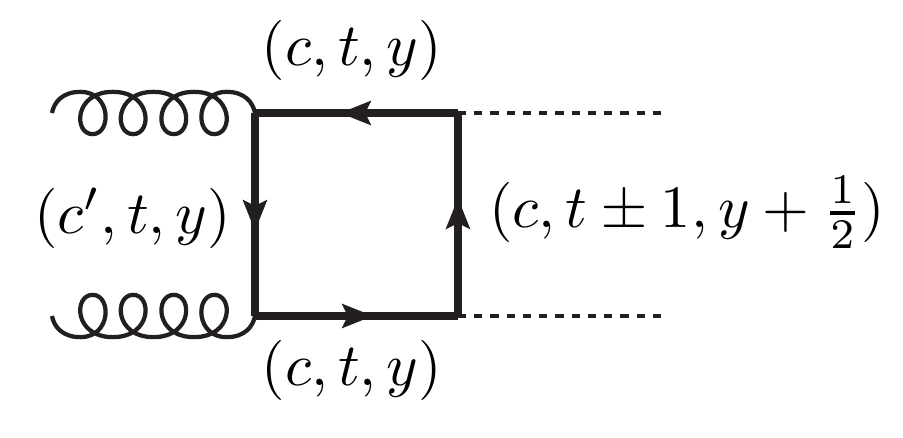}
\caption{Diagram leading to the operator $\mathcal{O}_{HG}$ with the particles in the loop labeled by their transformations under $SU(3)\times SU(2)\times U(1)$, i.e., $(c,T,Y)$ if $\bar{c}\otimes c'\ni 8$. If the particles in the loop are bosons, additional diagrams can be obtained by replacing one or two internal lines and their two adjacent vertices by a single vertex. }
\label{fig:OHG}
\end{figure}
\begin{figure}[b]
	\centering
		\includegraphics[width=0.45\textwidth]{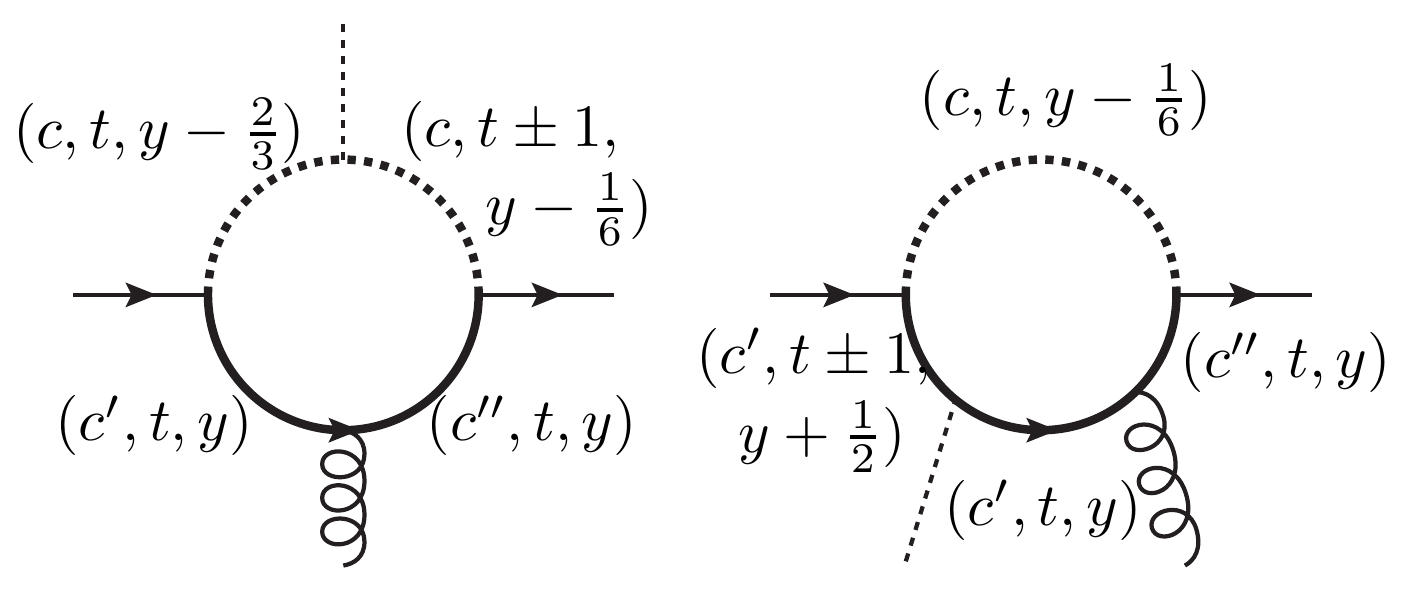}
\caption{Diagrams leading to the operator $\mathcal{O}_{hg}$ if $\bar{c'}\otimes c''\ni 8$, $\bar{c}\otimes c'\ni 3$ and $\bar{c}\otimes c''\ni 3$. The internal fermion and boson lines can be exchanged and the internal bosons do not have to be scalar. Similarly as for Fig.~\ref{fig:OHG}, additional diagrams can be obtained by removing one internal boson propagator. }
\label{fig:chromo}
\end{figure}

Taking $m_t=174.3$ GeV, $m_H=125$ GeV, $v=246$ GeV, $\Lambda=1$ TeV and $g_s = 1.2$, we obtain
\begin{equation}
\delta c_{HG}\approx0.1 \Re{c_{hg}}-0.006 c_y.
\end{equation}
Even if the effects due to the new interactions of the top quark are loop suppressed, they cannot be neglected. The coefficient $c_y$, probing the relation between the top mass and its Yukawa coupling, is not constrained by any other process than Higgs production (see recent and rather weak constraints on $c=1-c_y (v/\Lambda)^2 $ in Refs.~\cite{Azatov:2012bz,Espinosa:2012ir}). Similarly, the present constraints on $c_{hg}$ due to top pair production~\cite{Degrande:2010kt} including the latest ATLAS combination~
\cite{atlascombi}, 
i.e., $-0.75\lesssim c_{hg} (\mbox{TeV}/\Lambda)^2 \lesssim 3$ at 1 $\sigma$ and $-1.2\lesssim c_{hg} (\mbox{TeV}/\Lambda)^2 \lesssim 3.5$ at 2 $\sigma$, still allow the contribution from the chromomagnetic operator to have a noticeable effect on the allowed range for $c_{HG}$ as will be illustrated in the summary plots of Sec.~\ref{tth}. 

The next question concerns the typical expectation for the size of the coefficient $c_{HG}$.
For example, the one-loop contributions 
from $\mathcal{O}_{H}$ and $\mathcal{O}_{Hy}$ have been shown to be as large as the $\mathcal{O}_{HG}$ contribution in little Higgs models~\cite{Low:2010mr}. The reason is that those operators $\mathcal{O}_{H}$ and $\mathcal{O}_{Hy}$ can be induced by the tree-level exchange of a heavy particle while $\mathcal{O}_{HG}$ is only generated at the loop-level in a perturbative UV completion of the SM (see Fig.~\ref{fig:OHG}). The operator $\mathcal{O}_{H}$ is also enhanced compared to $\mathcal{O}_{HG}$ in strongly interacting Higgs models~\cite{Giudice:2007fh}.
%

On the contrary, the chromomagnetic operator can hardly be enhanced. It is also generated only at the loop-level (see Fig.~\ref{fig:chromo}) in perturbation theory and thus for $\mathcal{O}_{hg}$ to be the dominant new physics effects  requires $\mathcal{O}_{HG}$ to be relatively suppressed. While the diagram of Fig.~\ref{fig:OHG} can be obtained by using twice the lower part of the second diagram in Fig.~\ref{fig:chromo}, the first diagram in Fig.~\ref{fig:chromo} with $c=1$ does not imply the presence of $\mathcal{O}_{HG}$. As a consequence, it is possible to generate the chromomagnetic operator, ${\cal O}_{hg}$,  at one-loop and not the operator $\mathcal{O}_{HG}$.
An explicit example is given in Appendix~\ref{app:example}.
While dominant in this example, the effects from the chromomagnetic operator are too small to be observed. Alternatively, the hierarchy may come from strongly coupled theories and can be estimated with the help 
of Naive Dimensional Analysis~\cite{Georgi:1992dw, Manohar:1983md}. If only the right-handed top is strongly coupled, the dominant operator involves four top quarks yet does not contribute even at two-loop~\cite{Nomura:2009tw}. In that case, the coefficient of the chromomagnetic operator is only suppressed by one power of the strong coupling compared to two for $c_{HG}$ and both operators can have similar contribution when the strong coupling approaches $4\pi$. However, its effects may again be to small to be observed. So, let us now move to study the effect of these operators on $t\bar{t}h$ production.


\section{$t\bar{t}h$ production}\label{tth}

\begin{figure}[t]
\centering
		\includegraphics[width=0.35\textwidth]{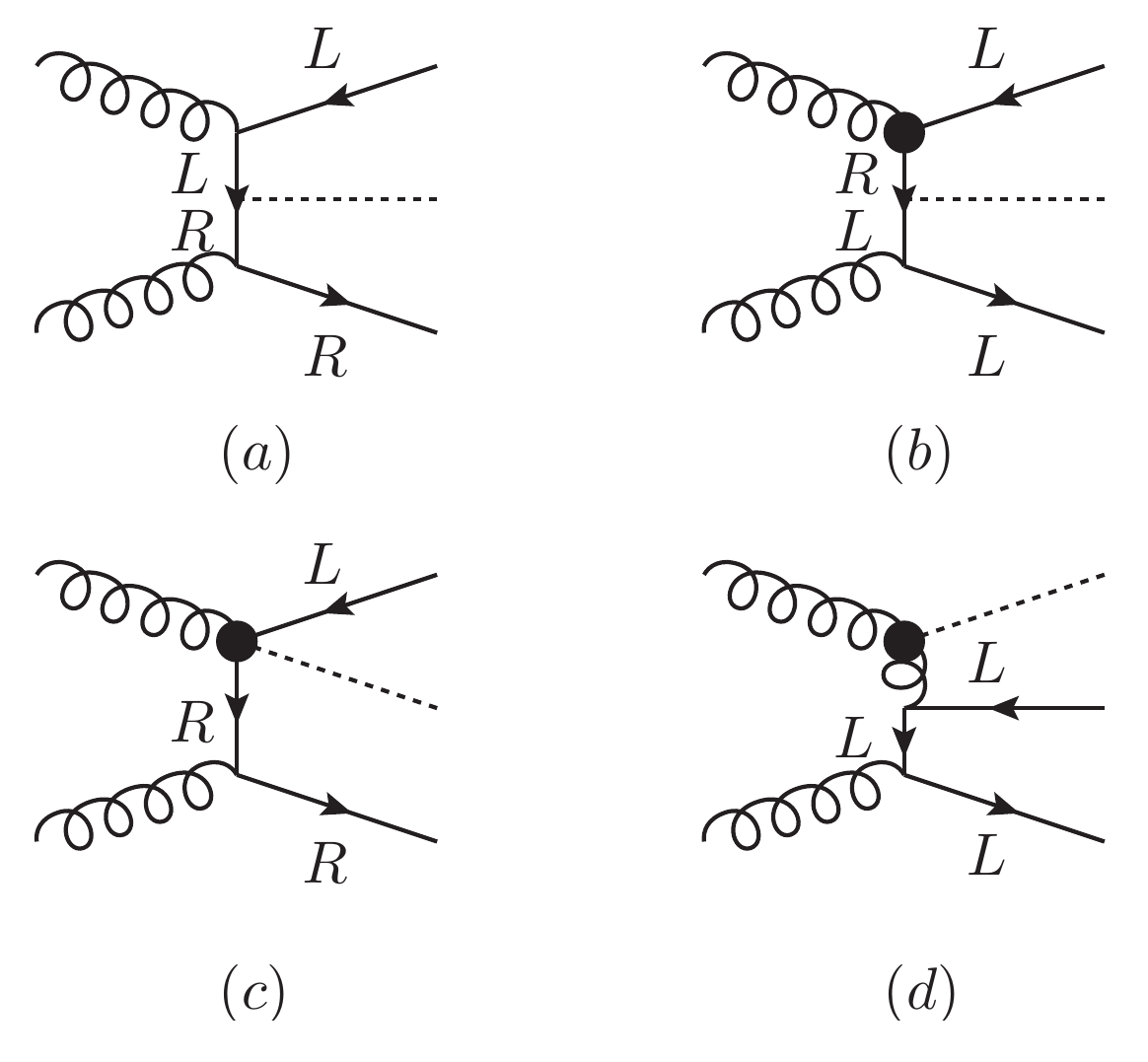}
\caption{Examples of diagrams for $t\bar{t}h$ production from the SM (a), from the chromomagnetic operator  $\mathcal{O}_{hg}$ (b) and (c) and from the $\mathcal{O}_{HG}$ operator (d). $\mathcal{O}_{H}$ and $\mathcal{O}_{Hy}$ lead to a simple rescaling of the SM contribution.}
\label{fig:ttHproddiag}
\end{figure}

While both Higgs direct coupling to the gluons and new top interactions significantly affect Higgs production, they cannot be distinguished using this process only. Contrary to Higgs production by gluon fusion, the four operators $\mathcal{O}_{HG}$, $\mathcal{O}_{hg}$, $\mathcal{O}_{H}$ and $\mathcal{O}_{Hy}$ all contribute to $ t\bar{t}h$ at the tree-level (see Fig.~\ref{fig:ttHproddiag}). Again, the three operators in Eq.~(\ref{htop}) have no contribution for this process due to parity. There is only one additional operator affecting this process, 
\begin{equation}
\mathcal{O}_G = f^{ABC}G_{\mu}^{A\nu}G_{\nu}^{B\rho}G_{\rho}^{C\mu}.
\end{equation}
However, this operator involves neither the top quark nor the Higgs
boson and is thus not expected to be enhanced. Moreover, the
interference between $O_G$ and the SM has a suppression similar to that
associated with the octet exchange  in top pair production
($\propto \beta^2 m_t^2$)\cite{Zhang:2010dr}. Indeed, the contribution
proportional to $c_G$ in the squared amplitude for $gg \to t\bar{t}h$
vanishes at threshold and becomes constant at large $s$. Large shape
effects on energy dependent distributions are thus not expected from
this operator. Consequently, although we include this operator in the calculation of the cross section, we 
do not consider it in our phenomenological analysis and set $c_G=0$ in all plots. 
The four-fermion operators cannot modify the main process, i.e., gluon fusion. Consequently, their contributions are about one order of magnitude smaller and have not been included.
%
%
%
The contribution from  $\mathcal{O}_{H}$ and $\mathcal{O}_{Hy}$, being just a rescaling of the top Yukawa coupling (see Eqs.~(\ref{eq:httren}) and (\ref{eq:cy})), is proportional to the SM cross section:
\begin{equation}
\sigma\left(pp\to t\bar{t} h\right)=\sigma\left(pp\to  t \bar{t} h\right)_{SM}\left(1- {c_y}\frac{v^2}{\Lambda^2}\right)^2
\end{equation} 
and this relation holds at NLO (at least in the flavor universal limit).
The total cross-section at 14 TeV is given by 
\begin{eqnarray}
\label{tth14}
\frac{\sigma\left(pp \to t\bar{t} h\right)}{\text{fb}} &=& 611^{+92}_{-110} +\left[ 457^{+127}_{-91} \Re{c_{hg}}- 49^{+15}_{-10}c_{G}\right.\nonumber\\
&+ & \left. 147^{+55}_{-32} {c_{HG}}- 67^{+23}_{-16} c_{y}\right]\left(\frac{\text{TeV}}{\Lambda}\right)^2 \nonumber\\
& + & \left[543^{+143}_{-123}  (\Re c_{hg})^2+1132^{+323}_{-232} c_{G}^2  \right.\nonumber\\
&+&\left. 85.5^{+73}_{-21} c_{HG}^2 +   2^{+0.7}_{-0.5} c_{y}^2    \right. \\
&+&\left. 233^{+81}_{-144}\Re{c_{hg}}c_{HG}   -50^{+16}_{-14} \Re{c_{hg}}c_{y}      \right.  \nonumber\\
&-&\left.   3.2^{+8}_{-8}\Re{c_{Hy}}c_{HG}- 1.2^{+8}_{-8} c_{H}c_{HG}
\right]
\left(\frac{\text{TeV}}{\Lambda}\right)^4  \ , \nonumber
\end{eqnarray}
%
%
\begin{figure}[t]
	\centering
		\includegraphics[width=0.5\textwidth]{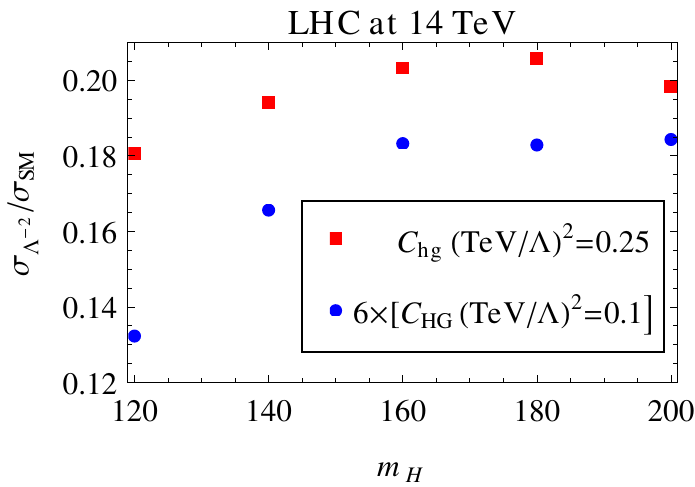}
		\caption{Ratio of the interference (between the SM and the main dimension-six operators) and the SM 
		 $p p \to t\bar{t} h$ cross section as a function of the Higgs mass. CTEQ6l1 pdf set and $\mu_R=\mu_F=m_t=174.3$ GeV are used and the results are very similar at 7 TeV. }
		\label{fig:thhmh}
\end{figure}
 %
at 8 TeV by 
\begin{eqnarray}
\label{tth8}
\frac{\sigma\left(pp \to t\bar{t} h\right)}{\text{fb}} &=& 128 +\left[ 94 \Re{c_{hg}} -9.7 c_{G} + 27 c_{HG} \right. \nonumber\\
&-&\left.15 c_{y}\right] \left(\frac{\text{TeV}}{\Lambda}\right)^2 
+ \left[ 53.9 (\Re c_{hg})^2+137 c_{G}^2 \right. \nonumber \\
&+& \left. 9.6
c_{HG}^2 +0.4 c_{y}^2 + 19.3\Re{c_{hg}}c_{HG}  \right. \nonumber \\
&-& \left.   9.6\Re{c_{hg}}c_{y} - 1.2
\Re{c_{Hy}}c_{HG} \right. \nonumber \\
&-& \left. 0.7 c_{H}c_{HG} \right]\left(\frac{\text{1
TeV}}{\Lambda}\right)^4 \ ,
\end{eqnarray}
and at 7 TeV by
\begin{eqnarray}
\label{tth7}
\frac{\sigma\left(pp \to t\bar{t} h\right)}{\text{fb}} &=& 86.3^{+10}_{-15} +\left[ 63^{+20}_{-14} \Re{c_{hg}}   + 22.3^{+8}_{-7} c_{HG}\right.\nonumber \\
&-& \left.10.2^{+4}_{-2.5} c_{y}- 5.6c_{G}\right]\left(\frac{\text{TeV}}{\Lambda}\right)^2  \\
&+& \left[ 43.6^{+17}_{-12} (\Re c_{hg})^2+78.3 c_{G}^2 + 8.6^{+1}_{-3} c_{HG}^2  \right.\nonumber\\
&+&\left. 0.3 c_{y}^2 + 21^{+6}_{-2} \Re{c_{hg}}c_{HG} - 7.2^{+1}_{-1.7}\Re{c_{hg}}c_{y}  \right.\nonumber\\
&-&\left.  1.5^{+1}_{-1} \Re{c_{Hy}}c_{HG}- 1.1^{+1}_{-1} c_{H}c_{HG}\right] \left(\frac{\text{TeV}}{\Lambda}\right)^4 \ ,
 \nonumber
\end{eqnarray}
for $m_H=125$ GeV. We included $c_G$ and $c_G^2$ terms  for indication (but not $c_G c_i$ terms),
however, as mentioned earlier, we will
 set $c_G=0$ in the rest of the analysis. 
\begin{figure*}[t]
\begin{center}
		\includegraphics[width=0.32\textwidth]{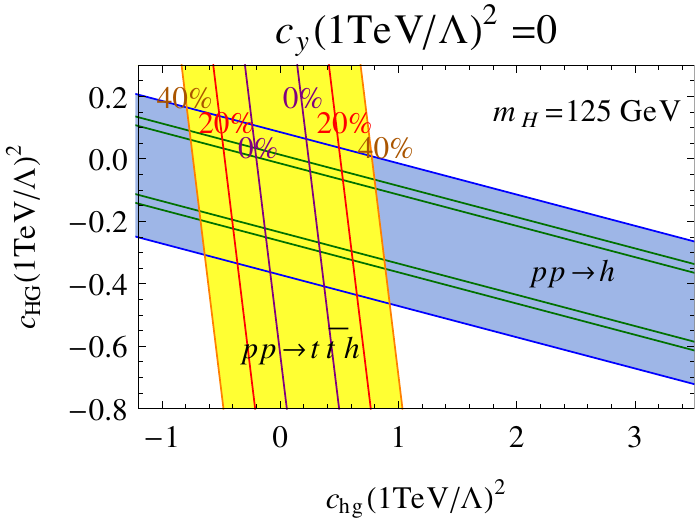}
		\includegraphics[width=0.32\textwidth]{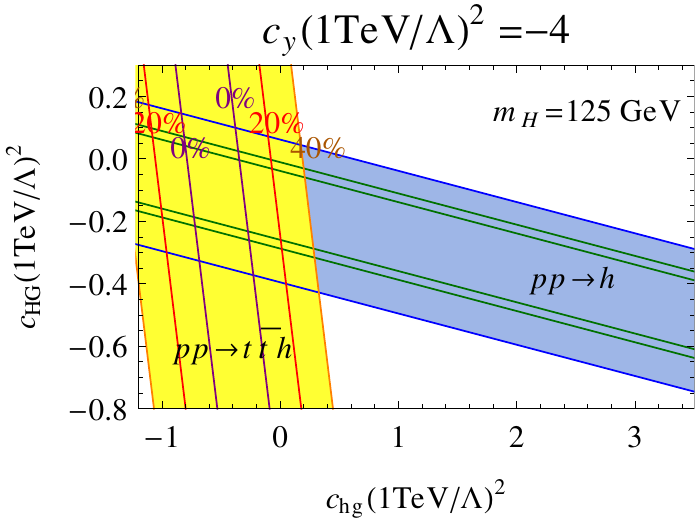}
		\includegraphics[width=0.32\textwidth]{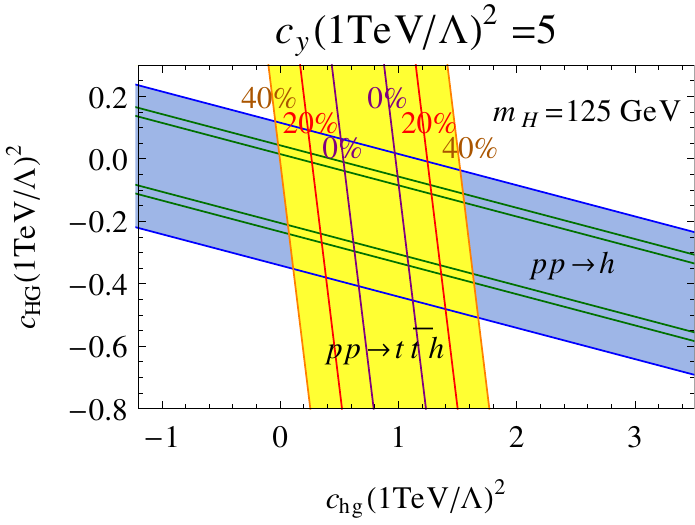}
\end{center}
\begin{center}		
		\includegraphics[width=0.32\textwidth]{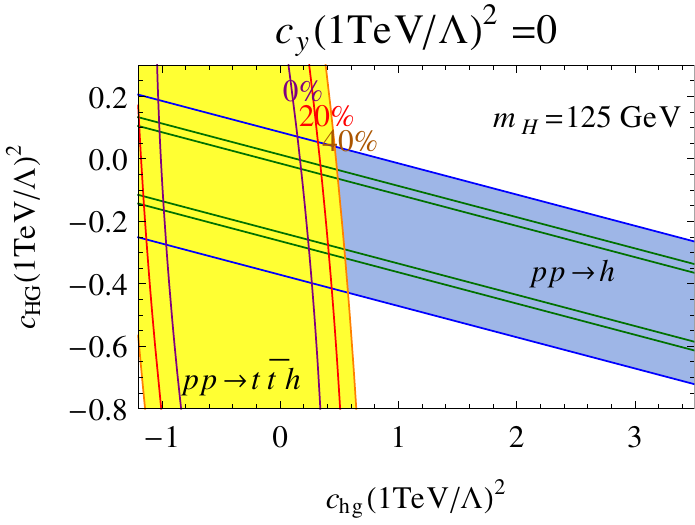}
		\includegraphics[width=0.32\textwidth]{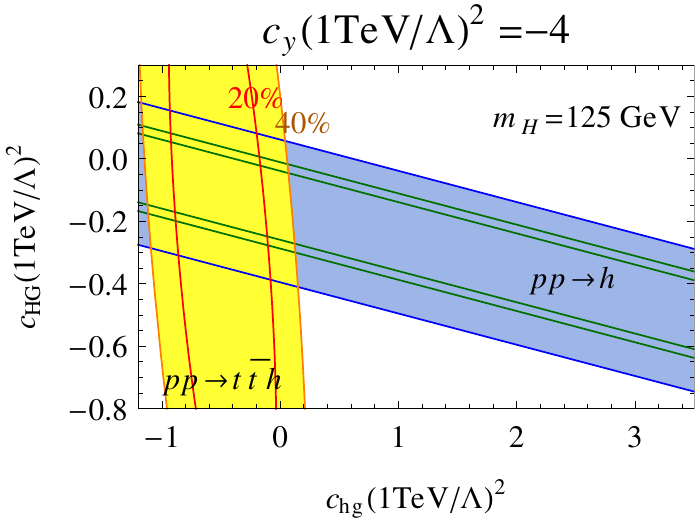}
		\includegraphics[width=0.32\textwidth]{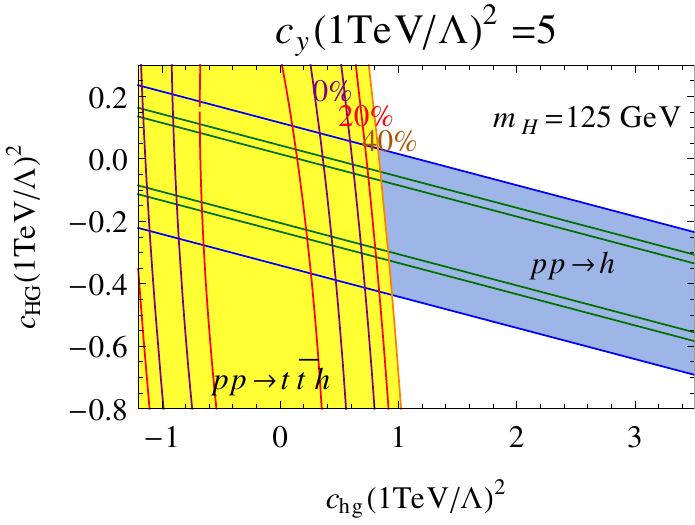}
\end{center}
		\caption{In blue, the region allowed by the Higgs production constraints at 7 TeV for $m_H=125$ GeV. 
		The green lines delimit the 2 allowed tiny bands obtained if the Higgs cross-section is measured
at its SM value with a precision of 20 $\%$.
		The yellow region is obtained by  assuming a 40\% precision on the $t\bar{t}h$ cross-section at 14 TeV with the measured central value matching the SM prediction and $c_G=0$.  The three plots correspond respectively to $c_y (\mbox{TeV}/\Lambda)^2=$0, -4, +5. The upper plots are obtained when neglecting the ${\cal O}(1/\Lambda^4)$  terms in the $t\bar{t}H$ cross section. The bottom plots instead include these higher order terms.  }
		\label{fig:combi}
\end{figure*}
%
\begin{figure*}[t]
  \centering  	
                 \includegraphics[width=0.45\textwidth]{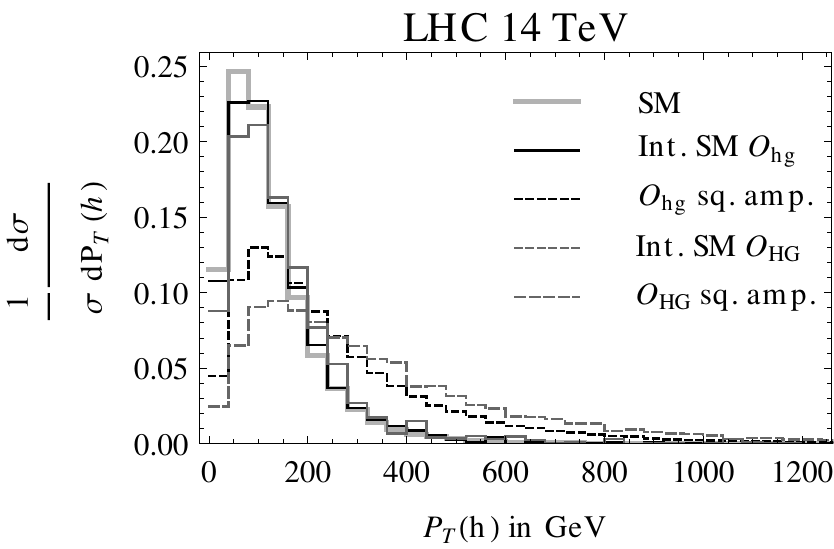}\hspace{.4cm}
                 \includegraphics[width=0.47\textwidth]{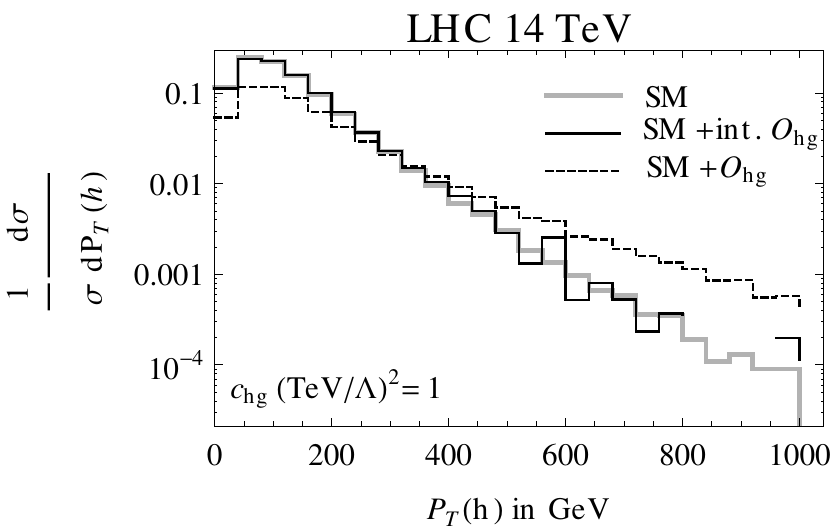}\\[.2cm]
		\includegraphics[width=0.45\textwidth]{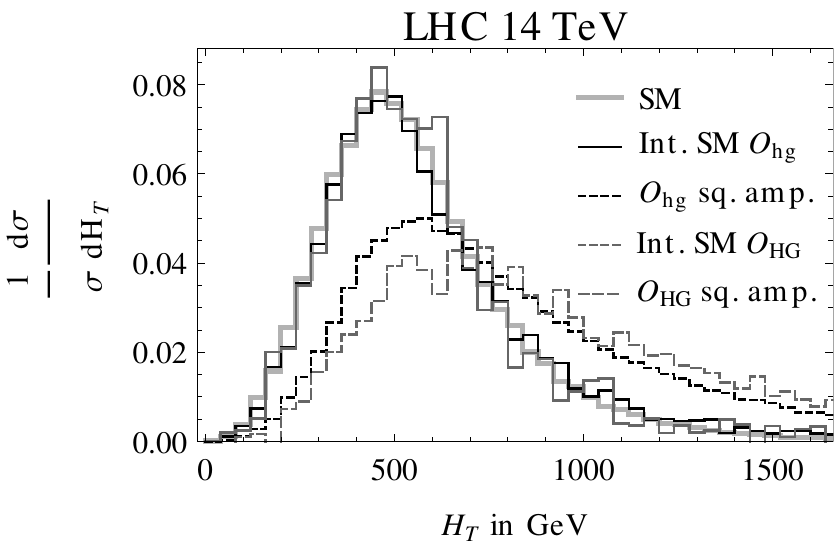}\hspace{.4cm}
		\includegraphics[width=0.47\textwidth]{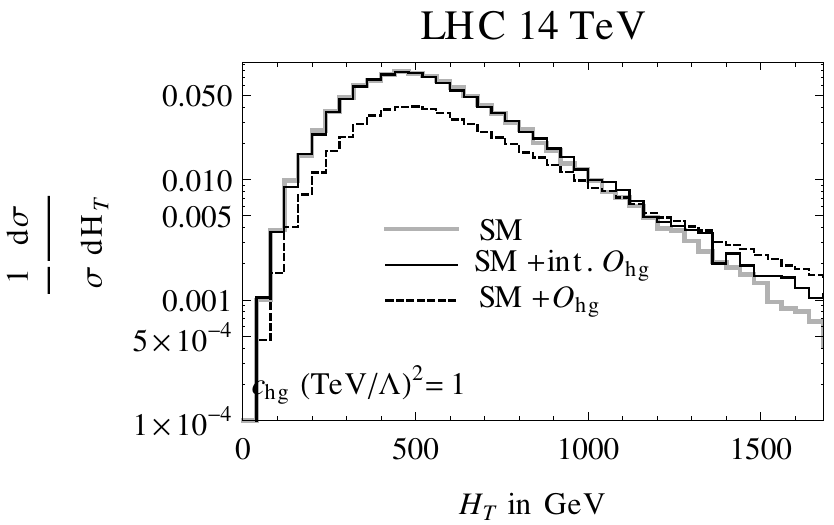}\\[.2cm]
		\includegraphics[width=0.45\textwidth]{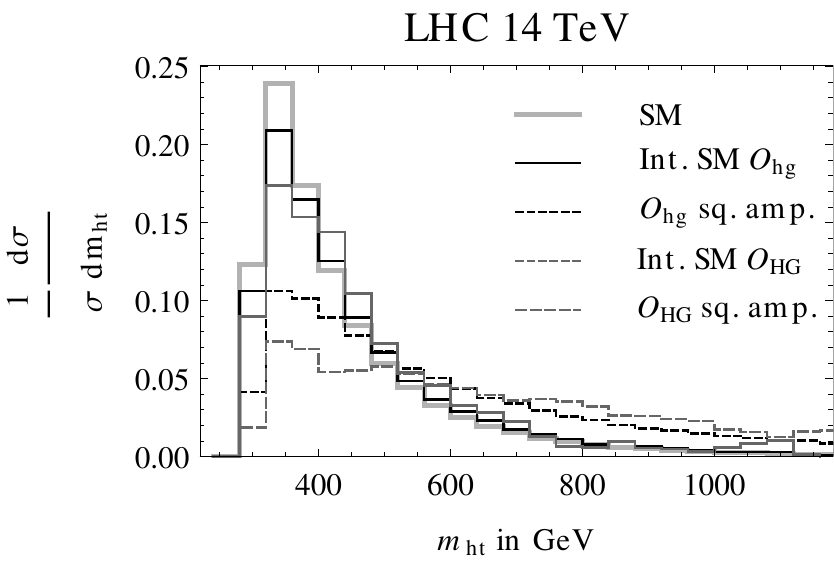}\hspace{.4cm}
		\includegraphics[width=0.47\textwidth]{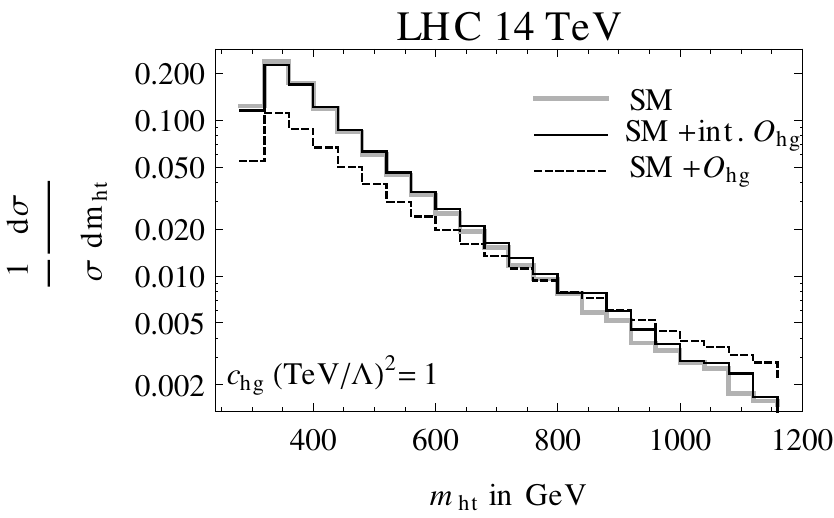}
						\caption{Left: Normalized distributions of the Higgs transverse momentum $P_T(h)$, the total $H_T$  and the invariant mass of the Higgs-top system using CTEQ6l1 pdf set, $\mu_R=\mu_F=m_t=174.3$ GeV and $m_H=125$ GeV for the SM, its interference with $\mathcal{O}_{hg}$ and $\mathcal{O}_{HG}$  and the squared of the amplitudes with one effective vertex. These plots do not depend on the value of $c_{hg}$ and $c_{HG}$. Right: 
						Total contribution (SM + $\mathcal{O}_{hg}$)  for $c_{hg} (\mbox{TeV}^2/\Lambda^2)=1$ including the interference terms only and including both the interference terms and the terms of order $1/\Lambda^4$, compared to the SM only. }
		\label{fig:Etth}
\end{figure*}
The same factorization and renormalization scales as for top pair production, i.e,  $\mu_F=\mu_R=m_t$ have been used since we have only considered a light Higgs boson. The cross-section will slightly decrease if higher values taking into account the Higgs mass are chosen. The errors are again obtained by varying the factorization and renormalization scales simultaneously from $\mu_F=\mu_R=m_t/2$ to $\mu_F=\mu_R=2m_t$, except for the last two
 terms $\Re(c_{Hy}) c_{HG}$ and $c_{H} c_{HG}$ for
which the numerical errors are larger. 
Results have been obtained via the FeynRules-MadGraph~5 simulation chain~\cite{Christensen:2008py,Degrande:2011ua,deAquino:2011ub,Alwall:2011uj}. The new physics has been computed at the tree-level and the  SM contribution at NLO~\cite{Dittmaier:2011ti,Beenakker:2001rj,Dawson:2002tg}. 
The ${\cal O} (1/\Lambda^4)$ terms have been computed to check the $1/\Lambda$
expansion and only take into account the  operators that contribute also at the
$1/\Lambda^2$  order, i.e.,
contain either squares of the operators  ${\cal O}(1/\Lambda^2)$ or the interference of the SM with an amplitude
involving two new vertices. Additional contributions from
the operators in Eq.~(\ref{htop}) or dimension-eight operators and proportional to
the imaginary part of $c_{Hy}$ or $c_{hg}$ are not included.  
The values of the $1/\Lambda^4$ coefficients tell us that the $1/\Lambda$ expansion breaks down around the TeV for $c_i=1$. This lower value compared to top pair production \cite{Degrande:2010kt} is expected due to the higher energy required for this final state.
While Eqs.~(\ref{tth14})$-$(\ref{tth7}) have been obtained only for a particular value of the Higgs mass, the ratios of the new physics contributions over the SM do not change drastically with the Higgs mass  as shown on Fig.~\ref{fig:thhmh}.

As shown by Eqs.~(\ref{tth14})$-$(\ref{tth7}) and Fig.~\ref{fig:thhmh}, $t\bar{t}$ associated Higgs production can mainly be affected by the chromomagnetic operator. As a consequence, the constraints from a measurement of the $t\bar{t}h$ cross-section would complement those from Higgs production  as illustrated in Fig.~\ref{fig:combi}, which displays the $c_{hg}$ range allowed by the present measurements of the  $t \bar{t}$ cross section.
By the time the $t\bar{t}h$ cross section will be  measured, the improved constraints from $t\bar{t}$ measurements will also help in reducing further the allowed range for $c_{hg}$ according to  Ref.~\cite{Degrande:2010kt}:
\begin{equation}
\left.\delta \sigma_{pp \to t \bar{t}} \right|_{14\mbox{ \tiny TeV}}= 144 \ c_{hg} \left(\frac{\mbox{TeV}}{\Lambda}\right)^2 + 22.5\  c^2_{hg}\left(\frac{\mbox{TeV}}{\Lambda}\right)^4.
\end{equation}

Like for top pair production, the theoretical uncertainty is responsible
for a sizable part of the allowed region. Since those errors mainly
affect the overall normalisation, this issue could be solved by
measuring the shapes of the distributions. Additionally, shape effects could
also lift the remaining degeneracy between the four operators.
While the contributions of the operators $\mathcal{O}_{Hy}$ and $\mathcal{O}_{H}$ have the same shapes as the SM ones, the operators $\mathcal{O}_{hg}$ and $\mathcal{O}_{HG}$ can induce shape distortions. However, only the contribution of the chromomagnetic operator might have a higher energy dependence than the SM. If the Higgs leg is attached to the effective vertex, the diagrams contain only one chirality flip such that no other chirality flip is needed to interfere with the SM amplitude (Fig.~\ref{fig:ttHproddiag}(c)). Moreover, the vertex is not proportional to the Higgs vev like for top pair production. Those advantages are lost if the Higgs is attached to the top line or to a gluon (Figs.~\ref{fig:ttHproddiag}(b) and (d)). For those diagrams, the amplitude is proportional to $m_t$ and $v$ and no room is left for extra powers of the energy of the process.

The distributions of the transverse momentum  of the Higgs, the total $H_T$ and  the invariant mass of the Higgs-top system are displayed on Fig.~\ref{fig:Etth}. The shapes of the $1/\Lambda^4$ contributions are also shown for comparison. They are clearly stretched to high energy while the interference and the SM contributions have a very similar behavior. The interference with the diagrams in which the Higgs is connected at the effective vertex do not vanish but are apparently suppressed. The shape effects are only expected if the new physics scale $\Lambda$
is close to the maximal energy probed because they are due only to the
$1/\Lambda^4$ contributions.
The plots on the right show how the distributions can differ with respect to the SM in the case $c_{hg} (\mbox{TeV}^2/\Lambda^2)=1$.

Finally, spin correlations could exhibit some  dependence on $c_{hg}$. In the case of $t\bar{t}$ production,
the deviations due to $c_{hg}$  were of the order of a few percents \cite{Degrande:2010kt}.
For $t\bar{t}h$, the measurement will be much more challenging and we therefore do not compute the associated spin correlations here but might return to them in due time. 

\section{Conclusion}

Only one dimension-six operator, $\mathcal{O}_{HG}$, generates a tree level coupling between the Higgs boson and the gluons. This operator has the largest contribution to Higgs production. Nevertheless, the three operators modifying the contribution from the top loop also have sizable effects compared to the SM one and, in a large class of models, can be  comparable to the effect of $\mathcal{O}_{HG}$ due to the hierarchy between their coefficients.
All those operators are already constrained by the present limits on Higgs production at hadron colliders. However, Higgs production by gluon fusion only constrains a linear combination of these operators and cannot discriminate between them. 
Interestingly, a light Higgs makes real the possibility of partially solving this issue by using 
Higgs production in association with a pair of top quarks. Contrary to Higgs production, the leading contribution in this process comes from the chromomagnetic operator $\mathcal{O}_{hg}$, which can therefore be further constrained from the measurement of the total $t \bar{t} h$ cross-section. Shape effects do not come from the interference terms and are dominated by the square of the amplitude involving an effective vertex and could thus be observable  for large $c_{hg}$ values only.\\

{\bf Acknowledgments:} 
We would like to thank C. Zhang for pointing out a mistake in our original computation of the one-loop corrections to $\mathcal{O}_{HG}$. The work of C.D., J.-M. G. and F.M. is supported by the Belgian Federal Office for Scientific, Technical and Cultural Affairs through the Interuniversity Attraction Pole No. P6/11. C.D. is a fellow of the Fonds National de la Recherche Scientifique and the Belgian American Education Foundation. G.S is supported by the ERC Starting Grant 204072. 
C.G. is partly supported by the European Commission under the ERC Advanced Grant 226371 MassTeV and the contract PITN-GA-2009-237920 UNILHC.
We thank D. Choudhury and P. Saha for useful discussions about the contribution of the chromomagnetic operator in $gg \to h$. 

\appendix

\section{Explicit example with $c_{HG}  \ll c_{hg}$}
\label{app:example}

In this appendix, we provide a toy model in which the diagrams of Fig.~\ref{fig:chromo}
 are generated while the diagram of Fig.~\ref{fig:OHG} is not.
The new sector is given by 
\begin{eqnarray}
 T_{L,R}&\sim&\left(3,1,Y\right)\nonumber\\
 \Phi&\sim&\left(1,2,Y-1/6\right)\nonumber\\
 S&\sim&\left(1,1,Y-2/3\right)
\end{eqnarray}
where $Y\neq 2/3$ to avoid the mixing of $T$ with the SM top and $Y\neq -1/3$ to avoid the mixing between $\Phi$ and the Higgs doublet. The extra piece of the Lagrangian is given by
\begin{eqnarray}
 \mathcal{L}^{NP}&=& i \bar{T}\cancel{D}T - M \bar{T}T -\kappa \left(\bar{T}_R Q_L \Phi + \bar{Q}_L T_R \Phi^\dagger\right)\nonumber\\
&&  -\beta \left(\bar{t}_R T_L S^\dagger + \bar{T}_L t_R S\right) + D_\mu S^\dagger D^\mu S - M_S^2 S^\dagger S\nonumber\\
&& + \lambda_1  \left(S^\dagger S\right)^2 + \lambda_2 S^\dagger S H^\dagger H  + \lambda_3 S^\dagger S \Phi^\dagger \Phi\nonumber\\
&& + D_\mu \Phi^\dagger D^\mu \Phi - M_\Phi^2 \Phi^\dagger \Phi + \lambda_4  \left(\Phi^\dagger \Phi\right)^2 \nonumber\\
&& + \lambda_5 \Phi^\dagger \Phi H^\dagger H + M_3 \left( \Phi^\dagger H^\dagger S +  H \Phi S^\dagger \right) \ ,
\end{eqnarray}
where the  parameters $M$, $M_S$, $M_\Phi$ and $M_3$ are around or above the TeV scale. The model has an accidental $Z_2$ symmetry under which all the SM model particles are even while the new ones are odd. This symmetry prevents any tree-level generation of the higher dimensional operators when the heavy particles are integrated out. The operator $\mathcal{O}_{HG}$ cannot be generated at one-loop since the colored particle does not couple to the Higgs. 
On the contrary, the equivalent operator for the photon cannot be avoided. Indeed, even if the fermions can be chosen to be neutral, all the new scalars cannot be simultaneously neutral. The constraints from gluon fusion in the low mass will change with the branching ratio to two photons.
Nevertheless, the chromomagnetic operator is induced at one-loop and its coefficient given by
\begin{eqnarray}
 \frac{c_{hg}}{\Lambda^2} &=& \frac{\kappa \beta g_s M_3 }{4\left(4\pi\right)^2M^3}\Bigg[\frac{R_\Phi^2 \left(1-3 R_S^2\right)+R_S^2+1}{\left(R_\Phi^2-1\right)^2 \left(R_S^2-1\right)^2}\nonumber\\
&&+\frac{4 \left(\frac{R_S^4 \log
   (R_S)}{\left(R_S^2-1\right)^3}-\frac{R_\Phi^4 \log (R_\Phi)}{\left(R_\Phi^2-1\right)^3}\right)}{R_\Phi^2-R_S^2}\Bigg]
\end{eqnarray}
where $R_S\equiv\frac{M_S}{M}$ and $R_\Phi\equiv\frac{M_\Phi}{M}$. 

\bibliographystyle{JHEP.bst}

\end{document}